\documentclass{elsart3p}

\usepackage{natbib}

\usepackage{graphicx}

\usepackage{amssymb}

\usepackage{color}

\newcommand{\be}{\begin{equation}}
\newcommand{\ee}{\end{equation}}
\newcommand{\ba}{\begin{eqnarray}}
\newcommand{\ea}{\end{eqnarray}}
\newcommand{\brr}{\begin{array}}

\newcommand{\err}{\end{array}}
\newcommand{\bc}{\begin{center}}
\newcommand{\ec}{\end{center}}

\begin{document}

\begin{frontmatter}



\title{Dark matter accretion wakes of high-redshift black holes}


\author[label1]{Roya Mohayaee},
\author[label2]{Jacques Colin}

\address[label1]{Institut d'Astrophysique de Paris (IAP), CNRS, UMR 7095, 
UPMC, 98bis Bd Arago, 75014 Paris, France}
\address[label2]{Universit\'e Nice Sophia Antipolis, CNRS, Observatoire de la
C\^ote d'Azur, UMR 6202 (Cassiop\'ee), Nice, France}

\begin{abstract}

Anisotropic emission of gravitational waves during the merger or formation 
of black holes can lead to the ejection of 
these black holes from their host galaxies.
A recoiled black hole which moves on an almost radial bound 
orbit outside the virial radius of its central galaxy, in the cold dark matter background, 
reaches its apapsis in a finite time. 
The low value of dark matter velocity dispersion at high redshifts and the
black hole velocity near the apapsis passage yield a high-density wake
around these black holes.
Gamma-ray emission can result from the enhancement of dark matter annihilation in
these wakes. The diffuse high-energy gamma-ray background from the ensemble 
of such black holes in the Hubble volume is also evaluated.

\end{abstract}


\end{frontmatter}

\section{Introduction}
\label{sect:intro}

Accretion onto a black hole (BH) from a non-dissipative medium, such as
a dark matter (DM) dominated Universe, 
is rather inefficient, mainly due to the absence of a cooling mechanism. 
However, dark matter distribution can be highly modified in the presence of
a black hole.
Enhancement of dark matter density and {\it spike formation} in
the process of adiabatic accretion of dark matter
onto black holes has been studied extensively (Gondolo \& Silk 1999).
Here, we study a different mechanism and study the response of dark
matter to the BHs that are ejected from their host galaxies and move
in a cold dark matter background on bound orbits (Mohayaee, Colin \& Silk 2008).
Indeed, in this article we review in more detail the results already presented
in a short letter previously (Mohayaee, Colin \& Silk 2008).

Assuming a zero velocity dispersion for DM and a relative motion between 
the BH and DM, and assuming the flow in the frame of the 
BH is steady and uniform at infinity, the density diverges
on the downstream symmetry axis. This comes about because 
rings of dark matter, concentric with the symmetry
axis, shrink down to points on the axis and this loss of dimension manifests
itself in a singularity, i.e. a {\it caustic}. 
This caustic is merely a focal
line and forms due to axial symmetry and it is not a stable catastrophe 
as classified by Arnol'd (Arnol'd 1986).
Unlike a shock which forms in a dissipative 
medium, the particles cannot be trapped in a caustic 
and hence a caustic has very little mass in spite of its high density. 
However, due to the finite primordial dark matter velocity
dispersion, a high density wake, rather than a line singularity, forms. 
In this case some of the dark matter particles with velocities smaller than
escape velocities move on bound orbits around the BH instead of arriving from 
infinity on hyperbolic trajectories.

If dark matter consisted of weakly-interacting particles, such as those
which arise in supersymmetry or extra-dimension extensions of the
standard electroweak model, they would annihilate in pair and produce a host
of particles among which are very high-energy gamma-rays. The flux of
these gamma-rays depends on local dark matter density. Hence, the enhanced
density of the wake leads to an ever-more enhanced flux of high-energy
gamma-rays. It is this flux which we aim to determine in this work.

\begin{figure}
\begin{center}
\includegraphics[width=\columnwidth]{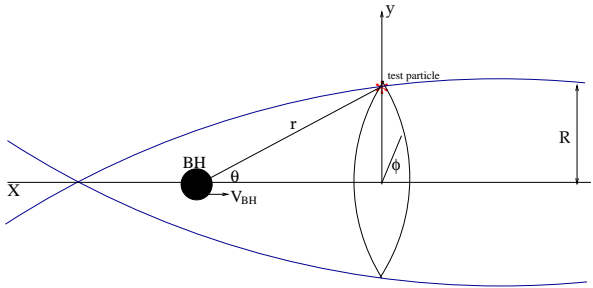}
\caption{The trajectory of a test particle in a cold (zero velocity dispersion)
medium
  past a perturber, {\it e.g.} a black hole.}
\label{fig:xfig-2008}
\end{center}
\end{figure}


How can black holes attain a velocity relative to the DM environment ? And can the velocity
dispersion of the background dark matter and the BH velocity be small enough for the wakes'
over-density to be significantly high ?

Black holes (BHs) can gain recoil velocity during their formation if 
their pregenitor stars collapse asymmetrically (Bekenstein 1973) 
or they can have
gravitational recoil during their mergers with other BHs (Fitchett 1983). 
The kick velocity can be as large
as a few hundreds of km/s during formation (Bekenstein 1973) and has recently been
shown to reach a few thousands of km/s for maximally
rotating BHs, with a particular spin configuration, during the merger phase (Campanelli et al 2007) . 

In spite of the recoil velocity, the density enhancement in 
the wake can be extremely low, because 
the velocity dispersion of dark matter is usually
very high. As dark matter collapses into a halo which then evolves 
by accretion and merger towards a final virialised state, its {\it effective} velocity 
dispersion increases. Our galactic halo is assumed
to have a velocity dispersion of a few hundreds of km/s.
This can be compared with the present primordial velocity dispersion of
dark matter which is only a few cm/s for neutralinos 
(that of axions is 7 orders of magnitude smaller).
However, as we go back in time, dark matter becomes less and less clumpy and
although the primordial velocity dispersion of dark matter increases linearly
with redshift (its present value is about 0.03 cm/s for neutralinos), 
its average velocity dispersion in the clumps falls as on average they become less massive and
smaller. Due to this low velocity
dispersion, many of the BHs with a moderate kick velocity
can escape from their host haloes at high redshifts (Favata et al 2004, Merritt et al
2004, Portegies Zwart 2000). The recoil velocity 
does not depend on the masses of the merging
BHs but only on the ratio of their masses and on
the spin configurations. Hence, at high redshifts
where dark matter haloes and escape velocities from them are small, only
a small recoil velocity is sufficient to set a BH on its orbit outside the
virial radius of its host halo.
In addition, at high-redshifts the velocity dispersion of dark matter is
mostly small and hence the wakes could be significantly overdense if the
BH also has a small velocity. The latter inevitably happens near the
apapses of the radial orbits.

The physics of the early Universe is unveiling fast (Barkana
\& Loeb 2001). To explain the quasar population at high redshifts, 
a large number of early massive BHs must have existed (see {\it e.g.} Madau \&
Rees 2001). These BHs are believed to have formed in 
dark matter haloes and grown by merger with other BHs and hence could 
attain recoil velocities.
These BHs can travel beyond the virial radii of their host
haloes and move through colder and colder environments.
At the apapses passages these BHs come to rest and there 
a substantial density
enhancement can occur. Dark matter annihilation in these high density regions 
could open a new window into the 
early Universe for high-energy gamma-ray explorers.

In Section \ref{sec:density}, we review the previous results on 
the calculation of the wake density and obtain the density profile in the 
limit as the velocity of the BH goes to zero.
In Section 3, we evaluate the time duration a BH spends 
around the apapsis of its orbit. 
In Section 4, we evaluate the absolute luminosity of a BH in high-energy
gamma-ray, as
a function of its mass and redshift. We also evaluate the boost factor
due to the BH wake relative to the background and also relative to the luminosity of 
the parent halo assuming a NFW profile. 
In Section 5, we evaluate the diffuse background, using
Press-Schechter formalism and compare our results with the minimum flux
from the host haloes. In Section 6, we conclude.

\begin{figure*}
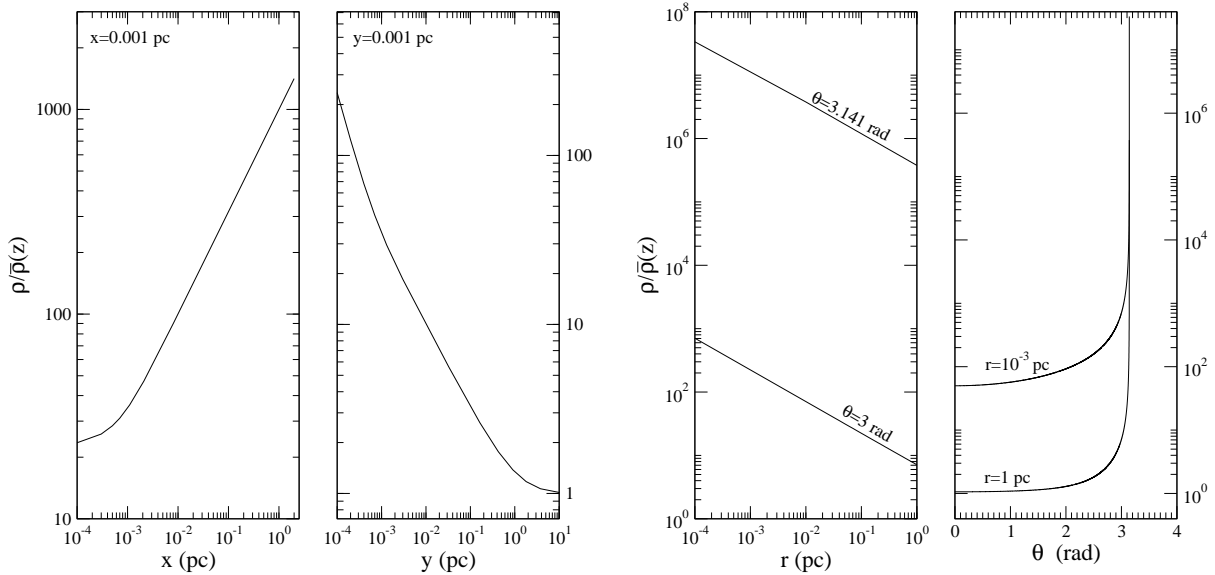

\begin{center}
\includegraphics[width=\columnwidth]{density-xy}
\hspace*{0.2cm}
\includegraphics[width=\columnwidth]{density-rtheta}
\caption{The density profile of dark-matter (with vanishing thermal
 velocity) around a BH of mass $M=100 M_\odot$ moving with a velocity of
 $1$ km/s. The coordinates 
are shown in Fig.~\ref{fig:xfig-2008} and Fig.~\ref{fig:wake-xfig}.}
\label{fig:density-xy}
\end{center}
\end{figure*}


\section{Density of the wake}
\label{sec:density}
We consider the black hole to be a point mass moving in 
a cold and almost homogeneous dark matter fluid. Due to this motion a wake forms 
behind the moving object which has a enhanced density w.r.t. the background.
Indeed, the enhanced density of the wake has been suggested as the underlying
reason for dynamical friction (Chandrasekhar 1949, Kalnaj 1971, H\'enon 1973). In this section we
evaluate the wake density for the three following cases: (I) a medium with zero velocity
dispersion, (II) when black hole velocity is comparable or smaller than the
velocity dispersion of its environment and (III) when the BH velocity is far larger than the
velocity dispersion of the dark matter medium.

\subsection{zero velocity dispersion: $\sigma_{\rm DM}=0$}

In the case of zero velocity dispersion, the density of the wake can be
evaluated by solving a two-body problem (Binney \& Tremain 1987). 
The orbit in polar coordinates (as shown 
in Fig.~\ref{fig:xfig-2008}) is given by
\be
{1\over r}={GM_{\rm BH}\over R^2 V_{\rm BH}^2}(1-{\rm cos}\theta)+{1\over R}{\rm sin}\theta\,.
\label{eq:orbit}
\ee

\noindent
where R is the impact parameter as shown in Fig.~\ref{fig:xfig-2008}. 
In this case, the 
density profile can be obtained analytically, using mass-conservation
and can be written in the following convenient form
\be
\left({\rho\over \bar\rho}\right)_{|\sigma=0}=
{1\over \sqrt{1-\left(1+{rV_{\rm BH}^2(1+{\rm cos}(\theta))\over
2GM_{\rm BH}}\right)^{-2}}}\;,
\label{density-cold-simple}
\ee
where $\bar\rho$ is the background density,
$M_{\rm BH}$ is the mass of the BH moving with velocity $V_{\rm BH}$ and the 
distance $r$ and angle $\theta$ are as 
shown in Figures \ref{fig:xfig-2008} and \ref{fig:wake-xfig}. Evidently 
a singularity forms on the axis behind the
BH for $\theta=\pi$ where the density diverges (see Fig.~\ref{fig:density-xy}). 
The density can also be written in terms of the cartesian coordinates as,
\be
\left({\rho\over \bar\rho}\right)_{|\sigma=0}={1\over 2}\left(\xi+{1\over \xi}\right)\;,
\label{density-cold-xy}
\ee
where
\be
\xi=\sqrt{1+{4\,G\,M_{\rm BH}\over V_{\rm BH}^2}\,\, {x+\sqrt{(x^2+y^2)}\over y^2}}\;.
\ee
and the profile is shown in Fig.~\ref{fig:density-xy}, where the
coordinates are shown in Figs.~\ref{fig:xfig-2008} and \ref{fig:wake-xfig}.
The situation considered here is unrealistic and dark matter particles
do indeed have a finite primordial velocity dispersion. This case is studied in
the next subsection.

\subsection{Density of the wake: $\sigma_{\rm DM}\not=0$}

\begin{figure}
\includegraphics[width=\columnwidth]{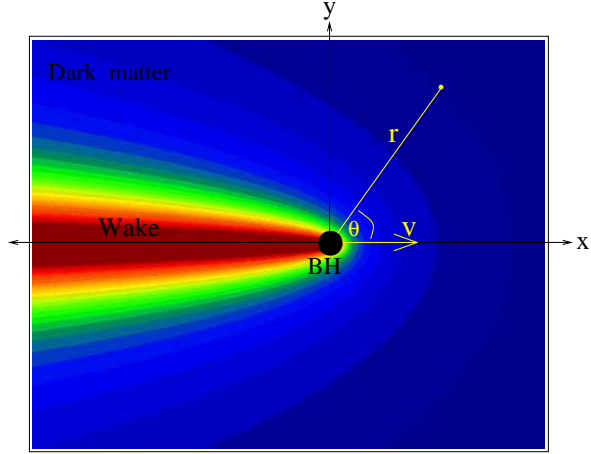}
\caption
{
Contour plot of the density field around 
a moving BH. The plots are made using
equation (\ref{eq:density-sweatman}) for a $10^3$ BH moving with a velocity of 1 km/s
in a dark matter field with velocity dispersion of 1 cm/s, 
and after a suitable normalization for graphical presentation. {\it
 colour coding: red-to-blue represents higher density to lower density.}
}
\label{fig:wake-xfig}
\end{figure}

In this case, we assume that the DM particles 
have non-zero temperature and their velocities obey Maxwellian
distribution.
The integral form of the density profile can be found by using Jeans' theorem.
The calculation is too detailed to be reviewed 
here (Danby \& Camm 1957, Griest 1988).
The final expression for the density
enhancement due to a moving point mass in a thermal environment is
\be
{\rho\over\bar\rho}=\int_{u=0}^{\infty}
du\,{u\sqrt{u^2+q^2}\over (2\pi)^{3/2}}\int_{\lambda=0}^\pi {\rm
  sin}\lambda\,d\lambda\int_{\nu=0}^{2\pi} d\nu\, e^{-F/2}\;,
\label{eq:density-griest}
\ee
where
\ba
F&=&p^2+u^2\\
&+ &2pu{(uZ+q^2({\rm cos}\theta)/2-\,Z\sqrt{u^2+q^2}{\rm cos}\lambda)\over
(u^2+q^2/2-u{\rm cos}\lambda\sqrt{u^2+q^2})}\;,\nonumber
\ea
and $p=V_{\rm BH}/\sigma_{\rm DM}$, $q=2GM_{\rm BH}/r/\sigma_{\rm DM}^2$, 
$Z=\sqrt{u^2+q^2}({\rm cos}\theta{\rm cos}
\lambda-{\rm sin}\theta{\rm sin}\lambda{\rm sin}\nu)$,
$\bar\rho$ is the density of the environment of the BH, and the 
distance $r$ and angle $\theta$ are as shown in Fig.~\ref{fig:xfig-2008}
and Fig.~\ref{fig:wake-xfig}.

\subsection{Density of the wake: $V_{\rm BH} \gg \sigma_{\rm DM}$}

When the velocity dispersion of the medium is very low 
in the limit $V_{\rm BH} \gg \sigma_{\rm DM}$, the integral in (\ref{eq:density-griest})
becomes highly oscillatory and difficult to evaluate. 
The small velocity dispersion, $\sigma_{\rm DM}$, of dark matter renders
the wake density finite downstream along the symmetry axis. 
For a dark matter velocity given by $-{\bf W}=(u-V,v,w)$, the wake
moves to a new position, and the density enhancement is given by
\be
{\rho\over \bar\rho}=\int_u\!\!\int_v\!\!\int_w 
{ du\,dv\,dw\,\, f(u,v,w)\over 
\sqrt{1-\left(1+{W^2r(1-{\rm cos}(\psi))\over 2GM_{\rm BH}}\right)^{-2}}}\;,
\\
\label{density-hot-general}
\ee
where ${\rm cos}\phi={V-u\over W}, {\rm tan}\alpha={v\over w}$ and the
angles are related using spherical trigonometry relation
\be
{\rm cos}\psi=-{\rm cos}\theta{\rm cos}\phi+{\rm sin}\theta{\rm sin}\phi{\rm
  cos}\alpha
\ee
and subsequently
\ba
& &
W^2(1-{\rm cos}\psi)=\\
& &
W^2\left(1+{(V-u)/ W}{\rm cos}\theta-{w/ W}{\rm
  sin}\theta\right).\nonumber
\ea
Under the condition that $V_{\rm BH}\gg\sigma_{\rm DM}$, one can demonstrate, 
using expression (\ref{density-hot-general}) that away from the 
negative symmetry axis (downstream) the
density approaches that for zero velocity dispersion
(\ref{density-cold-simple}).
After a suitable 
re-arrangement of expression (\ref{density-hot-general}) one obtains (see Sweatman \&
Heggie 2004 and also Sikivie \& Wick 2002)
\ba
&  & \rho(M_{\rm BH},z,r,\theta)=
{2\bar\rho(z)V_{\rm BH}^2\over \pi\sigma_{\rm DM}^2(z)}\int_{n=0}^\infty\int_{\alpha=0}^\pi n\,dn 
\, d\alpha\,
\nonumber\\
&\times &
e^{-V_{\rm BH}^2n^2/\sigma_{\rm DM}^2(z)} {f(M_{\rm
BH},z,n,\alpha)\over\sqrt{f(M_{\rm BH},z,n,\alpha)^2-1}}\;,
\label{eq:density-sweatman}
\ea
where the function $f(M_{\rm BH},z,n,\alpha)$ is given by
\ba
f&=&
1+{rV_{\rm BH}^2\over 2GM_{\rm BH}}\\
& \times & \left(1+{\rm cos}\theta-2\sqrt{1+{\rm cos}\theta} \,
n\,{\rm cos}\alpha+n^2\right)\;.
\nonumber
\label{eq:f}
\ea
We integrate expression (\ref{eq:density-sweatman}) numerically and show the
density contours in Fig.~\ref{fig:wake-xfig}. However, 
one can show analytically that for $V_{\rm BH}\gg \sigma_{\rm DM}$, the 
density (\ref{eq:density-sweatman}) along 
the symmetry axis ($\theta=\pi$) attains the following maximum value:
\be
\rho_{|axis}\approx{\bar\rho(z)\over\sigma_{\rm DM}(z)}\sqrt{\pi\,G\,M_{\rm BH}\over r} \qquad {\rm for}
\qquad \theta=\pi 
\label{eq:densityaxis}
\ee
as long as $\sigma_{\rm DM}^2\ll GM_{\rm BH}/r$. 
That the wake density is independent of the velocity of the
BH, downstream along the symmetry axis, might be surprising. However, we recall that the
density is infinite there for a zero velocity dispersion and a cut-off to this
divergence is put by finite dark matter velocity dispersion.

The density enhancement along the symmetry axis can be 
obtained by direct integration of (\ref{eq:density-sweatman}).
The density increases with
squared-root of the mass of the black hole, grows quadratically with z and
can be shown to also fall slowly with distance (as $1/\sqrt{r}$). 
These results are confirmed by 
the simple expression (\ref{eq:densityaxis}).

In the case of zero velocity dispersion, the density enhancement given by 
(\ref{density-cold-simple}) has a $1/V_{\rm BH}$ dependence and hence diverges as
$V_{\rm BH}\rightarrow 0$.
For a non-vanishing velocity dispersion, the density enhancement given by the
integral (\ref{eq:density-sweatman}) is independent of 
velocity along the symmetry axis as shown by 
expression (\ref{eq:densityaxis}) (see Fig. \ref{fig:density-v}). 
Off the symmetry axis the density falls almost linearly with increasing
velocity, and flattens at large velocities as predicted by the exponential
term in expression (\ref{eq:density-sweatman}).

%
\subsection {Density of the wake : $V_{\rm BH}\rightarrow 0$}
%

\begin{figure*}
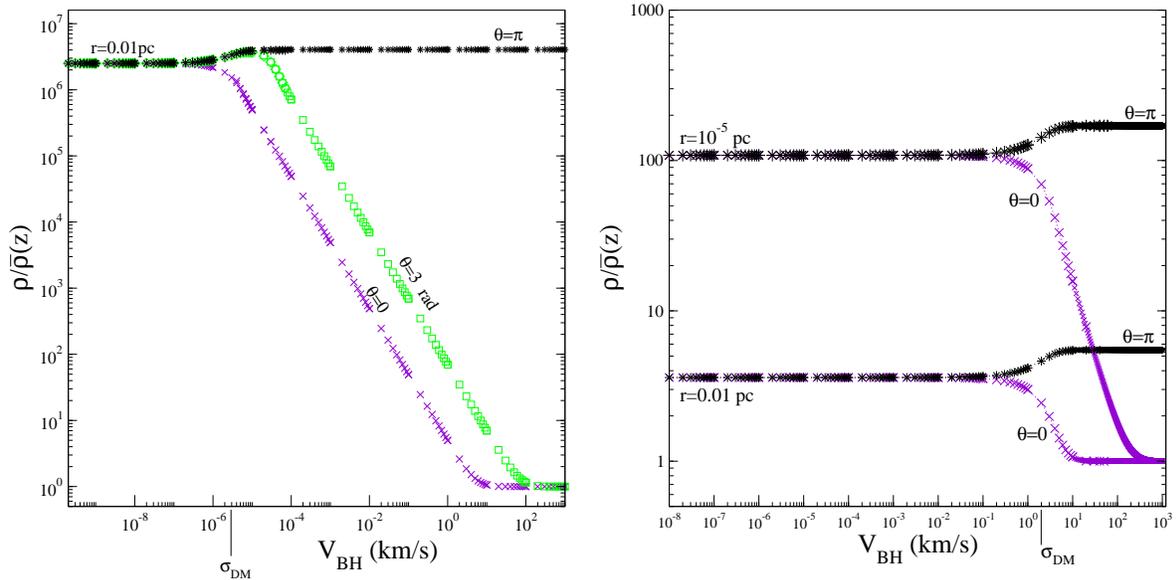

\includegraphics[width=\columnwidth]{density-v}
\includegraphics[width=\columnwidth]{density-v-hot}
\caption
{
The density (\ref{eq:density-griest}) for $V_{\rm BH}\le \sigma_{\rm DM}$
and (\ref{eq:density-sweatman}) for $V_{\rm BH}\gg \sigma_{\rm DM}$ are shown
for a $100 M_\odot$ BH moving in a cold ( of primordial velocity dispersion 
$\sigma_{\rm DM}=0.03 {\rm cm/s}$) [left panel] 
and hot  ($\sigma_{\rm DM}=2 {\rm km/s}$) [right panel] environment at
 high redshift (z=10). The
 right panel also demonstrates the $1/\sqrt r$ profile of the density.
}
\label{fig:density-v}
\end{figure*}

The zone of influence of BH decreases with increasing its velocity and
the velocity dispersion of its environment. 
In the limit as $V_{\rm BH}\rightarrow 0$, the
density profile of the wake (\ref{eq:density-griest}) reduces to
\be
{\rho\over \bar\rho}= \sqrt{4\over\pi}\sqrt{r_\bullet\over \,r}+
e^{r_\bullet/r}{\rm Erfc}\left(\sqrt{r_\bullet\over r}\right),
\label{eq:density-stationary}
\ee
where ${\rm Erfc}$ is the complementary error 
function and $r_\bullet$ is the radius of influence
of the BH: $r_\bullet=GM_{\rm BH}/\sigma_{\rm DM}^2$. We emphasis 
that (\ref{eq:density-stationary}) is the limit $V_{\rm BH}\rightarrow 0$ of
(\ref{eq:density-griest}), and is not a unique density profile for stationary
BHs. Here, we use (\ref{eq:density-stationary}) only as an approximation to
(\ref{eq:density-griest}) for slowly-moving BHs.
Fig.~\ref{fig:density-v}, shows the
dependence of the density enhancement on the BH velocity. 
When the BH is moving fast with respect to the background, a significant
density enhancement only arises in a small zone around the symmetry axis
(downstream) of the BH. The density enhancement also decreases with
increasing velocity dispersion of dark matter environment. 
The highest density enhancement and largest radius of influence are achieved for BHs
moving slowly ($V_{\rm BH}\le \sigma_{\rm DM}$) in a cold background.

Expressions (\ref{eq:density-griest}), or
(\ref{eq:density-stationary}), have been obtained by assuming that the BH
velocity remains constant, which is questionable in our situation where the BH is
both slowed down by the pull of its parent halo and also by the dynamical
friction of the wake itself. Since the BH is most luminous near the apapsis
where its velocity is very small, we expect our assumption to be indeed valid
near the apapsis. This is justified by Fig.~\ref{fig:density-v}: once
the BH velocity is less that the velocity dispersion of the background, the
constant velocity approximation is valid.


\section{Time spent around the apapsis}

The BH remains bound to its central halo if 
it is ejected with a velocity less 
than the escape velocity (measured from the virial radius). 
Because it is ejected from the centre (and also when with a large velocity),
the BH is on almost radial orbit.
The BH initial velocity is set as follows.
We assume that the halo mass is about $2\times\, 10^4$ times 
the mass of the BH (Madau \& Rees 2001), keeping in mind that the validity
of the Magorrian
relation (Magorrian et al 1998) at high redshifts is yet to be confirmed. Thus,
for a BH of mass $M_{\rm BH}$, the virial radius of the halo, from which it was
ejected, can be determined
using $M_{\rm halo}={4\pi/3}\Delta_{\rm vir}(z)\bar\rho(z)R_{\rm vir}^3(z)$
and noting that $\Delta_{\rm vir}(z)=(18\pi^2+82x-39x^2)/\Omega(z)$ and
$x=\Omega(z)-1$ and  
 $\Omega(z)=\Omega_m(1+z)^3/[\Omega_m(1+z)^3+\Omega_\Lambda+\Omega_k(1+z)^2]$
(see Bullock et al 2001 for details). Having evaluated the
virial radius, we can then evaluate the escape velocity from the
virial radius of the halo, using 
$V_{\rm escape}=\sqrt{2GM_{\rm halo}/R_{\rm vir}}$. 

The ejected BH is slowed down by the gravitational pull of its parent halo 
and also by the dynamical friction of 
dark matter background as
\be
\!\!{dV_{\rm BH}\over dt}\!=\!-\left[{(2\,E+V_{\rm BH}^2)^2\over 4\,G\,M_{\rm halo}}
+{4\,\pi\,G^2\,
M_{\rm BH}\bar\rho {\rm ln}(\Lambda) \over V_{\rm BH}^2} \right]
\label{eq:force}
\ee
where $E=-V_i^2/2+G\,M_{\rm halo}/R_{\rm vir}$ is the 
absolute value of the energy
with which a bound BH leaves the virial radius with velocity $V_i$.
Since dynamical friction plays a sub-dominant r\^ole in {\it braking} the BH,
the values of ${\rm ln}(\Lambda)$ and the background density 
$\bar\rho$ marginally affect the value of (\ref{eq:force}) for a BH in its
initial outward journey. 

By comparing the dynamical friction force to the force of the parent halo in (\ref{eq:force}), we
can find the range of values of the velocity for which the former dominates,
give by the inequality
\be
V_{\rm BH}(V_{\rm BH}^2+2\,E)\le\, 400 G M_{\rm BH}\sqrt{2\pi\bar\rho}
\ee
For a BH with ejection velocity (from virial radius) of about half the 
escape velocity, the inequality becomes
\be
V_{\rm BH}\le 4\,\times\,10^{-4}\left({M_{\rm BH}\over M_\odot}\right)^{1/3}\sqrt{1+z}
\ee
in km/s, for which the dynamical friction dominates over the pull of the parent halo.

Time spent at the apapsis is defined to be the time during which the velocity
of BH reduces from the background DM velocity dispersion
to zero (at the apapsis), i.e.  $0<  V_{\rm BH} < \sigma_{\rm DM}(z)$ 
where $\sigma(z)$ is the velocity dispersion of dark matter in the field
outside the halo. If
the dynamical friction dominates over the
gravitational pull of the halo, in bringing the BH to rest at apapsis. This yields 
\be
\Delta t_{\rm DF}\!=\!{V_{\rm BH}^3\over 12\pi G^2 M_{\rm BH} \bar\rho\,{\rm ln}(\Lambda)}\;,
\label{eq:tDF}
\ee
where Coulomb logarithm is set to unity here and is not expected to be
significantly greater than this value. 

If the pull of the halo is the dominant force in
bringing the BH to rest then the time during which the BH can be
considered stationary [hence its velocity $0<  V_{\rm BH} < \sigma_{\rm
DM}(z)$] is

\ba
& &\Delta t_{\rm halo}\!=\!{G\,M_{\rm halo}\over E}\left({\sigma_{\rm DM}\over
2E+\sigma_{\rm DM}^2}+{
{\rm Arctan}\left[{\sigma_{\rm DM}\over\sqrt {2E}}\right]\over \sqrt{2
E}}\right)\;.\nonumber
\\
\label{eq:tHP}
\ea

However, the density enhancement is most significant when the velocity of the
BH and the velocity dispersion of DM background are very low and hence when
the dynamical friction force dominates over the gravitational pull of
the host halo. Subsequently, $\Delta t_{\rm DF}$ is shorter 
than $\Delta t_{\rm halo}$.

The important issue we have not discussed here is that the wake does not form 
instantaneously and the time scale for the
formation of the wake has to be compared to the time the BH spends at the
apapsis. This would give us a meaningful estimate of the radius of the wake.
However, we postpone this issue to the forthcoming work and here leave
the radius of the wake, $R_{\rm
cutoff}$, as a free parameter. We estimate the suitable range of this
parameter which yields a sufficient luminosity. The criteria
here are that BH luminosity would dominate over that of its central halo and the
boost over the background luminosity evaluated within the same radius, $R_{
\rm cutoff}$, be far greater than unity.

\section{Gamma-ray flashes from BHs around apapses} 
\label{sec:Phi1BH}
Cold dark matter if composed of neutralinos or Kaluza-Klein particles
would annihilate in pairs and produce a host of secondary
products, including energetic photons (e.g. see Bertone, Hooper, Silk
2005 for a recent review).
The {\it absolute  luminosity}, in units of $\gamma {\rm s}^{-1}\,$ 
of a BH of mass $M$ at redshift $z$ is:
\be
L (M,z)\!\!=\!\!\left[{\cal N}_\gamma{\langle\sigma v\rangle\over 2m_\chi^2}\,\right]\!\!\!
\left[4\,\pi\int_{r_s}^{R_{\rm cutoff}}\!\!\!\!\! r^2 dr  
\left(\rho(M,z,r)\right)^2\,\right]
\label{eq:L}
\ee
where $r_s$ is the Schwarzchild radius, $r_\bullet$ is the radius of influence of the
BH, $m_\chi$ is the neutralino mass ($\sim 100$ Gev/c$^2$), 
$\langle\sigma v\rangle$ is the interaction cross-section [which we fix at  
$2\,\times\,10^{-26}$ cm$^3$/s] and ${\cal N}_\gamma$ is 
the number of photons produced per annihilation.
Note that the integral
(\ref{eq:L}) is independent of angle $\theta$ for stationary BHs.

We had previously found that the wake density of a slowly-moving BH ($V_{\rm
  BH}\le \sigma_{\rm DM}$) is well-approximated 
by the wake density of a stationary BH. 
By inserting (\ref{eq:density-stationary}) [keeping only the first
term] in (\ref{eq:L}) we obtain the analytic
expression (in units of $\gamma\, s^{-1}$)
\be
L_{\rm BH}=1.3\,\times\,10^{-15} \,(1+z)^6\,\,\left({M_{\rm BH}\over
  M_\odot}\right)\,{R_{\rm cutoff}^2\over \sigma_{\rm DM}^2}
\label{eq:LBH-new}
\ee
for the absolute luminosity of a BH
where $R_{\rm cutoff}$ and $\sigma_{\rm DM}$ are given in the same length units.
For dark matter with primordial velocity dispersion of $\sigma=0.03 (1+z)$
cm/s, expression (\ref{eq:LBH-new})
reduces to
\be
L_{\rm BH}=1.4\,\times\,10^{25}\,(1+z)^4\left({M_{\rm BH}\over
  M_\odot}\right)\,R_{\rm cutoff}^2
\label{eq:LBH}
\ee
for the absolute luminosity of a BH, where $R_{\rm cutoff}$ is in unit of
parsecs. We recall that in (\ref{eq:LBH-new}) and (\ref{eq:LBH}),
$R_{\rm cutoff}$ is the radius within which the luminosity of the BH is
evaluated. 

Next, the BH luminosity is compared to the background luminosity, $L_{\rm BG}$, within the
same radius in the absence of BH, which is given by
\be
L_{\rm BG}=5.36\,\times \,10^{-42}\,(1+z)^6\,R_{\rm cutoff}^3\;,
\label{eq:LBG}
\ee
where $R_{\rm cutoff}$ is in unit of cm.

\begin{figure}
\includegraphics[width=\columnwidth]{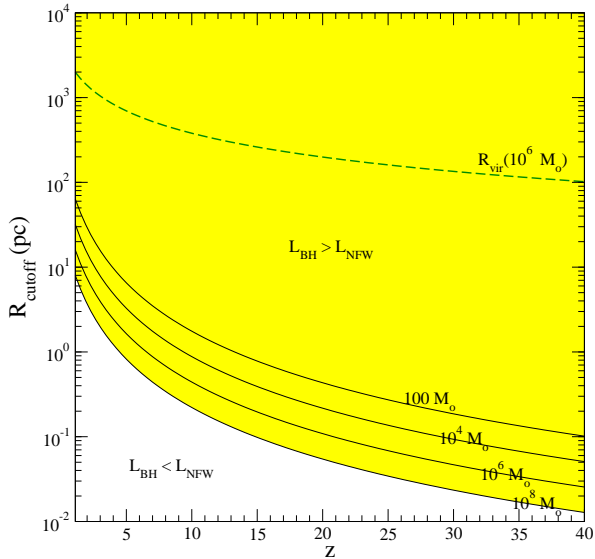}
\caption
{
The minimum radius $R_{\rm cutoff}$ at apapsis required for the BH luminosity to
dominate over that of its parent halo is shown by the lower solid lines
 for different masses.
 For reference, the 
minimum virial radius, $R_{\rm vir}$, of the halo (corresponding to ejection of a BH of mass
$100 M_\odot$) is also plotted.
}
\label{fig:rcutoff}
\end{figure}

The BH luminosity shall also be compared to the absolute luminosity of its central dark matter halo
 of mass $2\times 10^4\, M_{\rm BH}$, assuming it has a NFW density profile 
(Navarro, Frenk \& White 1997)
\be
{\rho\over \bar\rho_0}=(1+z)^3{\Omega_m\over \Omega(z)}{\delta_c\over
  [(c\,r/R_{\rm vir})(1+(c\,r/R_{\rm vir}))^2]}
\label{eq:NFW}
\ee
where 
$\Omega(z)=\Omega_m(1+z)^3/[\Omega_m(1+z)^3+\Omega_\Lambda+\Omega_k(1+z)^2]$,
$\delta _c=(\Delta_{\rm vir}/3) c^3/({\rm ln}(1+c)-c/(1+c))$,
 and $c$ is 
the concentration parameter, for which we use the fit
$[10/(1+z)] (M_{\rm vir}/M_{\*})^{-0.13}$ which agrees well with
Bullock et al 2001, Hennawi et al 2007 , and is slightly lower 
than Comerford \& Natarajan 2007. The halo mass within the virial radius is  
$M_{\rm halo}= M_{\rm vir}$.

The absolute luminosity  $L$ (\ref{eq:L}), in unit of $\gamma\,s^{-1}$,
of a NFW halo of mass $M_{\rm
halo}$ can then be evaluated  using (\ref{eq:NFW}) in (\ref{eq:L}) [after
setting $M=M_{\rm halo}, r_\bullet=R_{\rm vir}, r_s=0$] 
and fitted by the following functions
\ba
L_{\rm NFW} (z>1)  &=&  
{5.6\times 10^{27}\over (1+z)^{-1/3}}\left({M_{\rm halo}\over
  M_\odot}\right)^{0.7\,(1+z)^{0.075}}\nonumber\\
L_{\rm NFW} (z\le 1) &=& {1.6\times 10^{29}
\over (1+z)^{3}}\left({M_{\rm halo}\over
  M_\odot}\right)^{0.7}\,
\label{eq:LNFW}
\ea

A lower limit can be put on the parameter $R_{\rm cutoff}$ in (\ref{eq:LBH}), by requiring 
that $L_{\rm BH}/L_{\rm NFW}>1$. This is shown in Fig.
\ref{fig:rcutoff}. The figure shows that at high redshifts, very small
cutoff radii are sufficient for the
black holes to be more luminous than their host haloes. 

\begin{figure}
\includegraphics[width=\columnwidth]{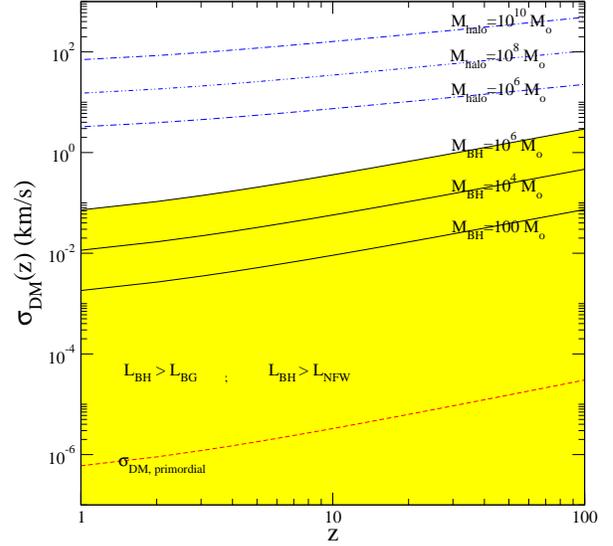}
\caption
{
The minimum value of the velocity dispersion required to make the BH
 more luminous than its central halo and also give a significant boost
 over the background, {\it i.e.} eq. (\ref{eq:boost}). The 
upper three dashed-dotted (blue) curves give the velocity dispersion of
 the parent halo, expression (\ref{eq:sigma-halo}), from which the BHs have been ejected. 
Clearly a very small velocity dispersion is required 
which is more likely to be satisfied at higher redshifts.
The bottom lower dashed (red) curve shows the evolution with redshift of the
primordial dark matter velocity dispersion.
}
\label{fig:sigma-z}
\end{figure}

However, in comparing the BH luminosity to that of its parent halo, we have
assumed a primordial velocity dispersion for DM. This assumption becomes less
realistic at lower redshifts. The velocity dispersion of dark matter
outside haloes has not yet been studied in numerical simulations. Observations
report on different values, depending on the environment. For
example in the local group, the velocity dispersion has been reported to be as
low as $40$ km/s (Karachentsev et al. 2003). Furthermore, it is not clear how
the velocity dispersion in the field evolves with
redshift. A high luminosity BH studied here, requires a low 
velocity dispersion environment. Here we put bound on this
velocity dispersion by requiring the BH luminosity to dominate over that
of the background and also over that of its central halo, {\it i.e.},
\be
L_{\rm BH}\gg L_{\rm BG} \qquad {\rm and} \qquad L_{\rm BH}\ge L_{\rm NFW}\;,
\label{eq:boost}
\ee
using expressions (\ref{eq:LBH-new}) for the black hole luminosity $L_{\rm BH}$,
(\ref{eq:LBG}) for the background 
luminosity $L_{\rm BG}$, and (\ref{eq:LNFW})
for the luminosity of the parent halo $L_{\rm NFW}$.  
The relationship (\ref{eq:boost}) puts an upper-bound on the velocity
dispersion.
The velocity dispersion of a dark matter halo
as a function of the redshift and the mass of its orbiting BH  
(recalling that $M_{\rm halo}=2\times 10^4 \,M_{\rm BH}$) is given by
\ba
\sigma_{\rm halo}\!&=&\!\left({M_{\rm BH}\over M_\odot}\right)^{1/3}\!\!\!\!\!\!\!\!\!\sqrt{(1+z)}
\left[12+{17\over 4\Omega(z)}-12\,\Omega(z)\right]^{1/6}\nonumber\\
&&
\label{eq:sigma-halo}
\ea
where again $\Omega(z)=0.3(1/(1+z)+0.3(1-1/(1+z))+0.7(1/(1+z)^3-1/(1+z)))^{-1}$.
The result is plotted in Fig.~\ref{fig:sigma-z}. In this figure, the lower three curves are
upper-bounds [see expression (\ref{eq:boost})] to
the velocity dispersion corresponding to BHs of masses $100 M_\odot, 10^4 M_\odot$ and $10^6
M_\odot$ as marked on the plot. The upper three dashed-dotted (blue) curves are plotted for
reference and represent the velocity dispersion of the dark matter haloes
from which the BHs were ejected, given by (\ref{eq:sigma-halo}).

Evidently the maximum velocity dispersion
outside the halo would be smaller than that inside the halo. 
Here we have given only an approximate outline and
we emphasis that a better method to clarify the problem of DM velocity dispersion
would be to
study the velocity dispersion outside DM haloes at 
different redshifts in N-body simulations. Although less decisive,
analytic studies can also be made through for example Press-Schechter formalism.

Further important constraint on $R_{\rm cutoff}$ can be put by studying the formation time
of the wake. Indeed, on one hand a finite time is required for a wake to form within a
certain radius and on the other hand the time spent at the apapsis
is not infinite. By comparing these two time scales, namely time for formation
of the wake and the time the BH spends at the apapsis (during which it can be considered
stationary) we can better determine the {\it radius of the wake}.
However, we do not expect this radius to be much smaller than
the minimum values of $R_{\rm cutoff}$ given in Fig.\ref{fig:rcutoff}. 
More detailed works on this problem is postponed to the forthcoming article and
here the wake radius is left as a free parameter.

\section{Diffuse gamma-ray background }
\label{sec:totalPhi}

Next, we study in a cosmological scenario, ejected BHs near their apapses passages, especially 
those at high redshifts, where the merger rate is
higher, the escape velocities are lower and the recoiled velocities are as
large as now. The recoil velocity depends on the mass ratio of the
BHs and not on the masses of the individuals, which indicates that ejected BHs
are expected to be more abundant at high redshifts.
Hence the ensemble of recoiled BHs might yield an 
observable diffused background flux. 
The total flux
is given by the integral
\be
\Phi=\int_M\int_z{L(M,z)\over 4\pi r(z)^2}\,N(M,z)\,dM\,d{\cal V}(z)
\label{eq:Phi}
\ee
where $M$ can be either the BH mass ($M_{\rm BH}$) or the halo
mass ($M_{\rm halo}$),
$r(z)=R_H\left(1-{1\over \sqrt{1+z}}\right)$ with Hubble radius
$R_H=4000 {\rm Mpc}$, and $N(M,z)$ is the number density of
the BHs [or haloes in the calculation for NFW haloes] 
and the luminosity of a single BH $L(M,z)$ (or the parent NFW haloes at z) is given 
by (\ref{eq:LBH}) [or (\ref{eq:LNFW})] and the volume element is 
$d{\cal V}={\sin}\psi\,r(z)^2 dr(z)  d\psi\,\,d\varphi$.

The physical number density of the BHs is assumed to follow the Press-Schechter
formalism (Press \& Schechter 1974, Bower 1991),
\ba
N(M,z)&=&{\bar\rho_0\over \sqrt{2\pi}}\left({n+3\over 3}\right)\left({M_{\rm halo}\over
  M_*(0)}\right)^{n+3\over 6}\,{(1+z)^4\over M_{\rm halo}^2}\nonumber\\
& &\times \exp\left[-{1\over 2}\left({M_{\rm halo}\over
	M_*(0)}\right)^{n+3\over 3}(1+z)^2\right] 
\label{eq:n}
\ea
where $n$ is
the power spectrum index $-2<n<-1$. For BHs, the above expression (\ref{eq:n}) has
to be 
multiplied by the {\it relative} time a BH spends at 
apapsis, {\it i.e.} $\Delta t_{\rm DF}/t_0$ where $t_0$ is the age of the Universe. 

The time spent at apapsis, (\ref{eq:tDF}), is itself a function of 
BH mass.
Thus only the fraction $\Delta t_{DF}/t_0$ of 
the ejected BHs can be considered to be
actually on their apapses passage in a Hubble time.
We can evaluate the
total flux
by performing the integrals in (\ref{eq:Phi}) 
[multiplied by $\Delta t_{\rm DF}/t_0$ for BHs]. In using
(\ref{eq:tDF}), we assume that the BH velocity $V_{\rm BH}$ is of the
same order as the dark matter velocity dispersion $\sigma_{\rm DM}$ whose
current value is $0.03$ cm/s and increases linearly with $(1+z)$. We
expect this assumption about the velocity dispersion to be more valid at
very high redshifts.
For spectral index $n=-1$ 
and $M_*\sim 10^{12}\,M_\odot$ and for the ensemble of BHs at their apapses
passages and their central haloes 
(assuming NFW profiles) we obtain 
$\Phi_{\rm NFW}\sim
10^{-6}  \qquad \gamma {\rm ~~ cm}^{-2}{\rm sr}^{-1}\,.
$
The flux from the BHs is lower than this value, since we have
evaluated our parameter $R_{\rm cutoff}$ by requiring $L_{\rm
  NFW}=L_{\rm BH}$. The BHs only spend a
fraction of time at the apapsis which yields
approximately
$\Phi_{\rm BH}\sim
10^{-14}  \qquad \gamma {\rm ~~ cm}^{-2}{\rm sr}^{-1}
$. \footnote {This corrects the typo in Mohayaee, Colin \& Silk 2008
  where this $\Phi_{\rm BH}$ was mistakingly written as $\Phi_{\rm NFW}$.}
The flux would be attenuated due to interaction of photons which however would affect
approximately equally $\Phi_{\rm BH}$  and $\Phi_{\rm NFW}$.
 
We have assumed that only BHs 
produced in $3\sigma$ peaks of the density perturbation can 
undergo effective mergers. 
We have assumed that all ejected BHs orbit their central haloes outside
the virial radius; 
however, were this not the case, we do
not expect any significant overall decrease in the flux which is already 
underestimated by our moderate choices of parameters 
 and also by assuming that there is
only one apapsis passage for a BH. We have also ignored the effect of multiple 
density-enhancement for a BH which is on its inward journey through an already
high-density wake.

\section{Conclusion}

Black holes can be ejected from their host haloes due to anisotropic
  emission of gravitational waves in the merger of their progenitors. If ejected with velocities below the
  escape velocity they move on bound
  orbits around their host haloes. Since they are ejected from the
  centres of the haloes they move on radial orbits and their velocities
  come to zero at the apapses passages. Around their apapses, these BHs
  have low velocities and move in a cold background and 
  very high density wakes can form around them. If dark matter was to consist of
  self-annihilating particles, these BHs would be powerful sources of
  high-energy $\gamma$-rays, both individually as resolved sources and collectively as
 diffuse background.
The results here indicate that
the globular clusters in the outskirt of our halo or field galaxies in our local 
Universe devoid of central BHs can have 
orbiting BHs which during their apapsis passages 
would produce flashes of high-energy $\gamma$-rays, although
this effect is expected to be most significant for massive objects at high redshifts.
The validity of dynamical friction formulae has been very rarely studied 
for radial orbits (Gualandris \& Merritt 2007). The fact that there is
 no mass loss makes BHs a rare case for
 dynamical friction theory. Throughout this work we have assumed a 
homogeneous background and a constant-velocity 
approximation, both of these assumptions are questionable for the problem
 considered here. 
The validity of these assumptions remains to be checked in high-resolution
  numerical simulations.

{\small 
We are grateful to Michel H\'enon, Joe Silk \& David Weinberg 
for contributions. We thank 
French ANR OTARIE for travel grants.}


\end{document}